\documentclass[preprint2]{aastex}
\usepackage{epstopdf}

\slugcomment{Not to appear in Nonlearned J., 45.}

\shorttitle{Gas-loss in Ursa Minor}
\shortauthors{Caproni et al.}

\begin{document}


\title{Three-dimensional hydrodynamical simulations of the supernovae-driven gas loss in the dwarf spheroidal galaxy Ursa Minor}


\author{A. Caproni, G. A. Lanfranchi and A. Luiz da Silva\altaffilmark{1}}
\affil{N\' ucleo de Astrof\' isica Te\' orica, Universidade Cruzeiro do Sul - Rua Galv\~ao Bueno 868, CEP 01506-000, S\~ao Paulo, Brazil}
\email{anderson.caproni@cruzeirodosul.edu.br}

\and

\author{D. Falceta-Gon\c{c}alves\altaffilmark{2}}
\affil{Escola de Artes, Ci\^encias e Humanidades, Universidade de S\~ao Paulo, Rua Arlindo Bettio 1000, CEP 03828-000 S\~ao Paulo, Brazil}




\altaffiltext{1}{present address: Observat\'orio Dietrich Schiel, Centro de Divulga\c{c}\~ao da Astronomia, Universidade de S\~ao Paulo, S\~ao Carlos, Brazil}
\altaffiltext{2}{present address: SUPA, School of Physics \& Astronomy, University of St Andrews, North Haugh, St Andrews, Fife KY16 9SS, UK}


\begin{abstract}
As is usual in dwarf spheroidal galaxies, today the Local Group galaxy \object{Ursa Minor} is depleted of its gas content. How this galaxy lost its gas is still a matter of debate. To study the history of gas loss in \object{Ursa Minor}, we conducted the first three-dimensional hydrodynamical simulations of this object, assuming that the gas loss was driven by galactic winds powered only by type II supernovae (SNe II). The initial gas setup and supernova (SN) rates used in our simulations are mainly constrained by the inferred star formation history and the observed velocity dispersion of \object{Ursa Minor}. After 3 Gyr of evolution, we found that the gas removal efficiency is higher when the SN rate is increased, and also when the initial mean gas density is lowered. The derived mass-loss rates are systematically higher in the central regions ($<300$ pc), even though such a relationship has not been strictly linear in time and in terms of the galactic radius. The filamentary structures induced by Rayleigh–-Taylor instabilities and the concentric shells related to the acoustic waves driven by SNe can account for the inferred mass losses from the simulations. Our results suggest that SNe II are able to transfer most of the gas from the central region outward to the galactic halo. However, other physical mechanisms must be considered in order to completely remove the gas at larger radii. 
\end{abstract}





\section{INTRODUCTION}

The Local Group is dynamically dominated by two spiral galaxies: Andromeda (M31) and the Milky Way (e.g, \citealt{mate98,berg99,berg07}). Less-luminous and less-massive galaxies reside around these systems (e.g., \citealp{mate98}).

Some of the least luminous and least massive galaxies in the local universe are dwarf spheroidal (dSph) galaxies, both classical (\object{Ursa Minor}, \object{Sextan}, \object{Draco}, \object{Fornax}, \object{Leo I}, \object{Leo II}, \object{Carina}, etc) and ultra-faint dwarf galaxies. The classical dwarfs present a total mass lower than about $10^8$ M$_\sun$ and an absolute $V--$magnitudes above $-14$, implying dynamical mass-to-light ratios larger than about 10 in solar units (e.g., \citealt{mate98,berg99,berg07,gre03,gre08,str07,wal13}). In contrast to dwarf irregular galaxies, dSphs are practically depleted of gas with \ion{H}{1} column densities smaller than $\sim 10^{19}$ cm$^{-2}$ \citep{youn99,youn00,grpu09}, dominated by old and intermediate-age stellar populations \citep{mate98,berg07}, even though some recent star formation activity has been inferred in some dSphs (e.g., \citealt{mate98,car02,lama04,dol05,bat06,berg07,kir11,boe12a,boe12b}).

Discovered by \citet{wils55}, the galaxy \object{Ursa Minor} is a good example of a classical dSph orbiting the Milky Way. Located at a heliocentric distance of about 64 kpc\footnote{There is a relative disagreement among the different estimates of the \object{Ursa Minor}'s distance in the literature (see the introduction of \citealt{pia05} and references therein for further information).} \citep{irha95}, \object{Ursa Minor} presents a spatial stellar distribution compatible with a King profile with a core radius of about 300 pc and a tidal radius roughly between 0.9 and 1.5 kpc \citep{irha95,kle98,pal03,str07}. Kinematic studies based on stellar velocity measurements show that \object{Ursa Minor} is a slow-rotating system (5$\pm$2 km s$^{-1}$; \citealt{mate98}) with a velocity dispersion of about 12 km s$^{-1}$ inside a radius of $\sim$36 arcmin from its nucleus \citep{wil04}. The integrated luminosity of \object{Ursa Minor} is about $(2-3)\times 10^5$ L$_\sun$, while its total mass ranges from 2 to $20\times 10^7$ M$_\sun$, implying a mass-to-luminosity ratio between about 70 and 800 \citep{irha95,kle98,gre03,str07,str08,wol10}.

The chemical properties of \object{Ursa Minor} are compatible with a relatively simple star formation history characterized by a single episode of star formation occurring from 13 to 10 Gyr ago \citep{lama04,lama07}. The lack of recent star formation activity agrees with the inferred upper limits of 430 and 7000 M$_\sun$ for the \ion{H}{1} mass \citep{youn00,gre03}, as well as the upper limit of $10^5$ M$_\sun$ for the \ion{H}{2} mass in \object{Ursa Minor} \citep{gal03}.

The question that arises is what mechanism(s) would be responsible for the gas removal in \object{Ursa Minor}. \citet{gre03} listed several physical mechanisms which could have removed the gas component in the context of dSphs. Supernova (SN) feedback is one potential gas loss channel. SN-driven galactic winds have been explored in the context of dSphs by several authors \citep{site98,mafe99,site01,fra03,mar06,sti07,rev09,qiwa12,rui13,recc14}, as well as in combination with external mechanisms such as ram-pressure and/or tidal stripping and ultraviolet (UV) background radiation (e.g., \citealt{may06,may07,saw10}). These works showed that supernovae (SNe) can heat and inject momentum into the interstellar medium (ISM), creating suitable conditions for the development of galactic winds. However, the efficiency of this process is not yet fully understood, mainly because of its dependence on the physical parameters of the galaxy (e.g., thermodynamical parameters of the gas, dark matter (DM) profile, etc.).

In this work, we study the process of gas removal in the specific case of \object{Ursa Minor} by taking into account observational estimates of its star formation history and by considering the effects of type II SNe (SNe II) feedback on the galactic mass loss. We aim to determine whether or not this process alone can explain the observed properties of the galaxy. \citet{lama03,lama04,lama07}, for example, were able to reproduce the chemical properties of this galaxy with a chemical evolution model characterized by a star formation history inferred by color-magnitude diagrams and assuming a galactic wind triggered by SNe. Their galactic wind, however, is established in an ad-hoc manner based only on the balance between the thermal and the binding energies of the gas. Several other different mass loss processes have been discussed in the literature (see Section 3.3 for further considerations on this), but we still lack a direct comparison to the chemical evolution and star formation histories of observed galaxies based on each of these, in particular for the case of \object{Ursa Minor}.

We run three-dimensional (3D) hydrodynamical (HD) simulations for a gas content in initial hydrostatic equilibrium with a static, cored DM gravitational field, taking into account the current knowledge about the star formation history of this object. Following \citet{lama07} and references therein, the simulations evolve the physical conditions of the gas during 3 Gyr, the interval in which stars were formed in \object{Ursa Minor}. 

This paper is structured as follows. In Section 2, we present an overview of the numerical code, as well as the initial conditions and numerical setup used in our 3D HD simulations in the context of the Local Group dwarf galaxy \object{Ursa Minor}. General results on the time evolution of the gas, the mass-loss efficiencies in terms of the SN II rates, and comparisons between our results and observational constraints are discussed in Section 3. The main conclusions obtained in this work are highlighted in Section 4.

\section{THE HD MODEL FOR URSA MINOR}

In this section, we provide a general overview of the numerical code, as well as the initial conditions and numerical setup used in the 3D HD simulations of the gas content of the dSph galaxy \object{Ursa Minor}.

\subsection{PLUTO Code: a Brief Overview}

PLUTO\footnote{\url{http://plutocode.ph.unito.it/}} is a finite-volume / finite-difference, shock-capturing code designed to integrate, in one, two, or three spatial dimensions, the differential equations \citep{mig07}:

\begin{equation} \label{conservlaw}
\frac{\partial \textbf{\emph{U}}}{\partial t} = - \nabla\cdot\mathbf{T}(\textbf{\emph{U}})+\textbf{\emph{S}}(\textbf{\emph{U}}),
\end{equation}
where $\textbf{\emph{U}}$ is the vector of the conservative quantities, $\mathbf{T}(\textbf{\emph{U}})$ is the flux tensor, and $\textbf{\emph{S}}(\textbf{\emph{U}})$ is the source term.

PLUTO can deal with the system of conservative laws in equation (\ref{conservlaw}) in classical HD, ideal/resistive magnetohydrodynamics problems, as well as in special relativistic hydrodynamics and ideal relativistic magnetohydrodynamics (e.g., \citealt{mig07,ros08,tes08,pofe10,sch10,bos12, bur12,mig12}).

In the HD case, including a radiative cooling function $F_\mathrm{c}$ and a gravitational potential $\Phi$ (see Section 2.2 for further information), $\textbf{\emph{U}}$, $\mathbf{T}(\textbf{\emph{U}})$ and $\textbf{\emph{S}}(\textbf{\emph{U}})$ are defined as (e.g, \citealt{mig07})

\begin{equation} \label{vectorU}
\textbf{\emph{U}} = \left(
\begin{array}{c}
\rho\\
\rho\textbf{\emph{v}}\\
E
\end{array}     \right)  ,
\end{equation}

\begin{equation} \label{tensorT}
\mathbf{T}(\textbf{\emph{U}}) = \left(
\begin{array}{c}
\rho\textbf{\emph{v}}\\
\rho\textbf{\emph{v}}\textbf{\emph{v}} + P\mathbf{I}\\
(E+P)\textbf{\emph{v}}
\end{array}     \right)^T  ,
\end{equation}
and

\begin{equation} \label{vectorS}
\textbf{\emph{S}}(\textbf{\emph{U}}) = \left(
\begin{array}{c}
0\\
-\rho\nabla\Phi\\
F_\mathrm{c} - \rho\textbf{\emph{v}}\cdot\nabla\Phi
\end{array}     \right)  ,
\end{equation}
where $\rho$ is the mass density, $P$ is the thermal pressure, $\textbf{\emph{v}}=(v_\mathrm{x}, v_\mathrm{y}, v_\mathrm{z})^T$ is the fluid velocity in Cartesian coordinates, and $\mathbf{I}$ is the identity tensor of rank 3. 

The total energy density $E$ is given by

\begin{equation} \label{totE}
E = \frac{P}{\Gamma-1} + \frac{\rho\vert\textbf{\emph{v}}\vert^2}{2},
\end{equation}
considering the ideal equation of state $P = (\Gamma-1)\rho\epsilon$, where $\Gamma$ is the adiabatic index of the plasma, assumed as 5/3 in this work, and $\epsilon \equiv E_\mathrm{int}/\rho = c_\mathrm{s}^2/[\Gamma(\Gamma-1)]$, since $P/\rho = c_\mathrm{s}^2/\Gamma$, where $E_\mathrm{int}$ and $c_\mathrm{s}$ are, respectively, the internal energy and the sound speed of the plasma.

The Cartesian grid of the computational domain is assumed to be fixed, i.e. invariant in time. Therefore, we do not take into account the effects of cosmological expansion in our simulations. Other aspects related to the cosmological evolution of the galaxy (e.g. interaction of the gas with cosmic background radiation, evolution of the DM potential, etc.) have also been neglected, as we consider them to be second-order processes.

\subsection{Initial Conditions and Numerical Setup}

\subsubsection{PLUTO Setup}

All 3D numerical simulations performed in this work were made assuming a cubic domain of 3$^3$ kpc$^3$, divided into a Cartesian grid of 256$^3$ points, where the HD equations were solved using the supercomputer Alphacrucis\footnote{Cluster SGI Altix ICE 8400. Further information on the Alphacrucis cluster at \url{https://lai.iag.usp.br}.} through the message passing interface (MPI) library for parallelization. The number of cores used to evolve our simulations in Alphacrucis varied from 256 to 520 (366 cores on average), which resulted in a total of about $9\times10^5$ processor hours to run the four main simulations discussed in this work.

Our HD numerical experiments evolved equation (\ref{conservlaw}) during an interval of 3 Gyr, the estimated duration of the star formation episodes in \object{Ursa Minor} \citep{lama04}.

\subsubsection{Boundary Conditions}

Before studying the impact of the SNe II on the ISM gas of \object{Ursa Minor}, we checked the numerical effects of the boundary conditions on the time distribution of the gas mass inside the computational domain. With this purpose, we run two additional simulations using the same initial gas and DM configuration as in our numerical simulations for Mgh05SN1 and Mgh05SN10 (see the next sections), but with different boundary conditions: (i) the standard {\it open} boundary conditions, in which the physical quantities at the ghost cells are equal to those in the boundary row of the domain but with gradients across the boundary set as zero; and (ii) an {\it outflow} boundary condition, in which outflow velocities are constant, but inflow fluxes, as well as gradients across the boundaries, are set as zero.

The initial gas distribution is let to evolve passively for 3 Gyr, without any internal or external perturbations (e.g., SNe, tidal forces, etc.). As the gas is in hydrostatic equilibrium with DM gravitational potential, its total mass is supposed to be the same in the end of the simulation, and the velocity field of the gas must remain null at all times. However, deviations from the perfect (initial) hydrostatic equilibrium due to the domain discretization (e.g., \citealt{zin02}) can deteriorate this scenario, introducing spurious mass fluxes between consecutive mesh cells.

In the case of the standard open boundary conditions, a large inflow flux of gas is observed with associated gas velocities reaching $\sim$250 km s$^{-1}$. These spurious fluxes produced an increment of  $1.4\times10^{13}$ M$_\sun$ inside a spherical radius of 950 pc (the present tidal radius of \object{Ursa Minor}). The open boundary condition acts as an infinite reservoir, which provides gas whenever the pressure equilibrium within the domain is broken. This spurious inflowing flux is a major issue for the assessment of the time evolution of the gas mass in \object{Ursa Minor}. 

The modified outflow boundary condition behaves in the opposite sense, with the total absence of a gas supply to increase the galactic mass. This boundary condition substantially reduced the spurious mass fluxes in the computational domain. The absolute values of the gas velocity were always below of 1 km s$^{-1}$. A mass of about $2.1\times10^6$ M$_\sun$ was added into a spherical region of 950 pc in radius after 3 Gyr of evolution, which means a small increase of 2.7\% in relation to the initial total mass of gas inside the same region. This small, but still present, influx of gas is the result of numerical errors associated with domain discretization and interpolation at the boundaries. The total mass inflow in this case is, however, several orders of magnitude smaller than that observed for the open boundary conditions, and small (few percent) compared to the initial total mass of the domain. Therefore, we set the boundary conditions for all models of \object{Ursa Minor} as outflow.

\subsubsection{The Cooling Function}

Radiative cooling for an optically thin gas is calculated in the PLUTO code through (e.g., \citealt{tes08})

\begin{equation} \label{cooling}
F_\mathrm{c}=\frac{\partial P}{\partial t} = - \left(\Gamma-1\right)n^2\Lambda(T),
\end{equation}
where $n$ is the number density of the gas and $\Lambda(T)$ is the cooling function.

In this work, we adopted the cooling rates derived by \citet{wie09}, which include additional effects of photo-ionization of heavy elements by UV and X-ray background radiation from galaxies (assuming the model of \citealt{hama01}), as well as by the cosmic microwave background radiation field. The cooling function used in our simulations was obtained from the interpolation of the precomputed tables provided by \citet{wie09}, assuming a gas number density of 0.1 cm$^{-3}$, roughly the initial mean number density inside the simulated box domain, [Fe/H] $\sim-2.13$, the median metallicity of \object{Ursa Minor} \citep{kir11}, and a radiation field found at a redshift of about 3.8. Note that this redshift corresponds to an age of the Universe of about $1.7$ Gyr, that is, roughly the half duration of the star formation in \object{Ursa Minor} derived by \citet{lama04} from chemical evolution models.

\begin{figure}
\epsscale{0.85}
\plotone{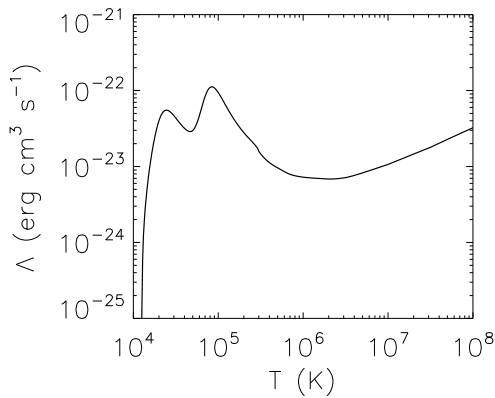}
\caption{Cooling rate function used in our numerical simulations of the gas loss in Ursa Minor.\label{coolfunc}}
\vspace{-0.4cm}
\end{figure}

In Figure \ref{coolfunc}, we show the cooling rate as a function of the temperature used in all of the numerical simulations performed in this work. For the sake of simplicity, this cooling function was used in all regions of the computational domain and during the whole computational time.

Radiative cooling was suppressed in regions with temperatures lower than $1.5\times 10^4$ K in order to mimic the equilibrium heating in the ISM of \object{Ursa Minor}, as well as to avoid the over cooling of the shocked gas produced by the SN blasts \citep{mafe99,fra03,fra04}.

\subsubsection{SN Rates}

Even though the SN rates in galaxies are variable over time, as well as across the galaxies themselves (e.g., \citealt{schm59,kenn98,recc14}), we have assumed in this work a constant rate in this work (in time and position) for the SN II explosions. 

We adopted two different time-constant SN rates, $R_\mathrm{SNII}$: 1 SN per Myr and 10 SNe per Myr. Both rates define the range of total SNe compatible with those expected from chemical evolution models for \object{Ursa Minor} along 3 Gyr of evolution \citep{lama04}. In addition, the star formation rates related to these SN rates can roughly generate a total stellar mass between 0.4 and 4$\times 10^6$ M$_\sun$\footnote{We have assumed a initial mass fuction of Salpeter for those calculations.} after 3 Gyr. This mass range is compatible with the inferred value of about 8$\times 10^5$ M$_\sun$ in stars for \object{Ursa Minor} \citep{dewo03,orb08}.

\subsubsection{Injection of Energy from SNe}

We assumed the following prescription to mimic the star formation rate, and consequently the occurrence of SNe II in our simulations: (i) we determined the SN rate timescale $t_\mathrm{R_\mathrm{SNII}} = R_\mathrm{SNII}^{-1}$; (ii) for each time step, we calculated the elapsed time $\Delta t$ between the occurrence of the last SN blast, $t_{i-1}^\mathrm{SNII}$, and the time $t$ for which calculations are being performed; (iii) when $\Delta t \geq t_\mathrm{R_\mathrm{SNII}}$, computational cells with number densities equal or superior to 0.1 cm$^{-3}$ are flagged as possible sites for SNe II take place. This is an usual ad hoc procedure in the literature (e.g, \citealt{katz92,summ93,kat96,kay02,rui13}) mainly because the impossibility of accessing molecular-cloud scales due to the numerical resolution of the simulations ($\sim$ 11.7 pc in our simulations); (iv) for each flagged cell, it is calculated the parameter $p_i$ (e.g., \citealt{katz92,moum06}), 
  
\begin{equation} \label{prob_sf}
p_i(t,x,y,z) = 1-\exp{\left[-C_*\frac{\Delta t}{t_\mathrm{ff}(x,y,z)}\right]},
\end{equation}
which measures the probability of an $i$-esime star formation episode occurring at the time $t$ and position $(x,y,z)$. The parameter $t_\mathrm{ff}$ is the free-fall timescale and $C_*$ is a dimensionless star formation rate parameter assumed to be 0.1 (e.g., \citealt{katz92,moum06}). Note that the SN rates in our simulations are insensitive to the adopted value of $C_*$, as well as to our own Equation (\ref{prob_sf}), since $R_\mathrm{SNII}$ is kept fixed throughout our simulations. This implies that $p_i$ is only used to assist in the choice of the SN sites; (v) We generated random numbers $p_\mathrm{aleat}$ between 0 and 1 for each flagged cell. If $p_\mathrm{aleat} \le p_i$, then the associated cell is chosen as a possible SN site \citep{moum06}. If there is more than one eligible site, then that with the highest value of $p_i$ is finally chosen to receive the SN event. (vi) Following \citet{fra04}, an internal energy of 10$^{51}$ erg is added to an approximately spherical volume with a two cell radius ($\sim$ 23.4 pc in our simulations) and centered at the elected SN site.

\subsubsection{DM and Initial Gas Density Profiles}

We have assumed that the gas component in \object{Ursa Minor} is under the influence of gravitational forces produced by a static, cored DM halo. This type of DM profile seems to be more suitable for dwarf galaxies \citep{burk95,sape97,bos00,blbo02,kle03,sim05,wal09,gov10,oh11,popo12,jage12}. The DM gravitational potential $\Phi_\mathrm{h}$ adopted in this work is mathematically defined as \citep{mafe99}

\begin{equation} \label{dm_pot} 
\Phi_\mathrm{h}(\xi) = v^2_\mathrm{c_\infty}\left[\frac{1}{2}\ln(1+\xi^2)+\frac{\arctan\xi}{\xi}\right],
\end{equation}
where  $\xi = r/r_0$ and $v_\mathrm{c_\infty}$ is the maximum circular velocity due to this DM potential:

\begin{equation} \label{vcircinf}
v_\mathrm{c_\infty}=\sqrt{4\pi G\rho_\mathrm{c}}r_0,
\end{equation}
where $G$ is the gravitational constant, and $\rho_\mathrm{c}$ and $r_0$ are, respectively, the central mass density and the characteristic radius of an isothermal, spherically symmetric DM mass density profile $\rho_\mathrm{h}$ \citep{bitr87,mafe99}:

\begin{equation} \label{dm_dens} 
\rho_\mathrm{h}(\xi) = \frac{\rho_\mathrm{c}}{1+\xi^2}.
\end{equation}

Following previous works (e.g., \citealt{mafe99,mar06,rui13}), we assumed an initial gas distribution in hydrostatic equilibrium with the DM gravitational potential. Under this condition, the initial mass density $\rho$ for an isothermal gas follows

\begin{equation} \label{rhogas} 
\rho(\xi) = \rho_0\exp{\left[-\Gamma\frac{v^2_\mathrm{c_\infty}}{c_\mathrm{s_0}^2}\chi(\xi)\right]},
\end{equation}
where $\rho_0$ is the initial mass density of the gas at the center of the galaxy, $c_\mathrm{s_0}$ is the initial sound speed, and $\chi(\xi)=v^{-2}_\mathrm{c_\infty}\Phi_\mathrm{h}(\xi)-1$.

The thermal gas pressure at iteration zero is calculated from

\begin{equation} \label{pgas} 
P(\xi) = \rho(\xi)\frac{c_\mathrm{s_0}^2}{\Gamma}.
\end{equation}

\begin{deluxetable}{cccc}
\tabletypesize{\scriptsize}
\tablecaption{Setup model parameters of our numerical simulations for the Ursa Minor galaxy.\label{setup}}
\tablewidth{0pt}
\tablehead{
\colhead{Model} & \colhead{$R_\mathrm{SNII}$} & \colhead{$M_\mathrm{g}$} & \colhead{$\rho_{0}$} \\
\colhead{} & \colhead{(Myr$^{-1}$)} & \colhead{(10$^7$ M$_\sun$)} & \colhead{(10$^{-23}$ g cm$^{-3}$)}
}
\startdata
Mgh19SN1 & 1 & 29.4 & 4.6\\
Mgh19SN10 & 10 & 29.4 & 4.6\\
Mgh05SN1 & 1 & 7.95  & 1.3\\
Mgh05SN10 & 10 & 7.95 & 1.3\\
\enddata
\tablecomments{Columns are as follow: (1) model identification; (2) SN II rate; (3) initial mass of the gas inside the tidal radius of \object{Ursa Minor}; (4) initial mass density at the center of the galaxy. All models assume $r_0=300$ pc, $\xi_\mathrm{t}=3.2$, $\xi_\mathrm{h}=50.0$, $c_\mathrm{s_0} = 11.5$ km$^{-1}$ (corresponding to a gas temperature of $\sim$9544 K), $v_\mathrm{c_\infty}=21.1$ km s$^{-1}$, $\rho_\mathrm{c}=6.2\times 10^{-24}$ g cm$^{-3}$ and $M_\mathrm{h}=1.51\times 10^9$ M$_\sun$ (see text for further details).}
\end{deluxetable}

\begin{figure}
\epsscale{1.0}
\plotone{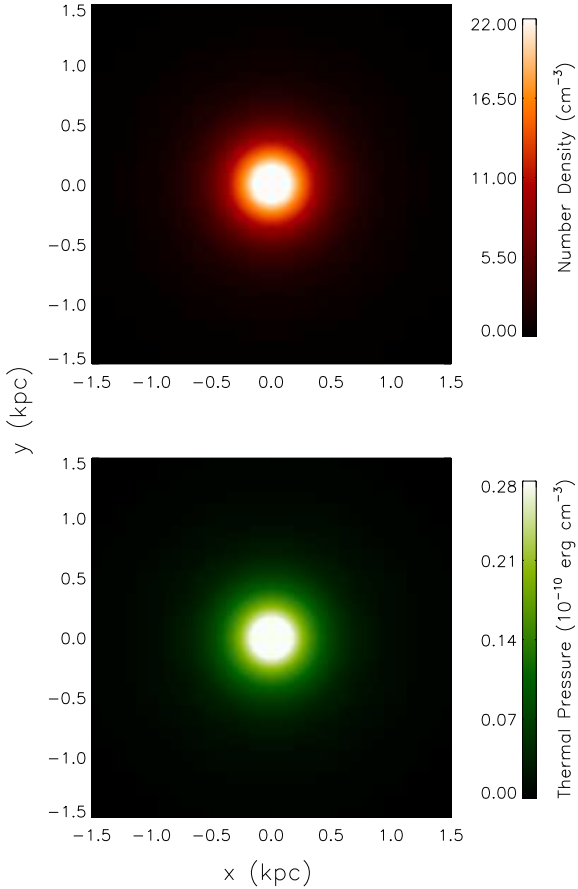}
\caption{Spatial distribution of the gas number density and pressure  (from top to bottom) on the $xy$ plane for the models Mgh19SN1 and Mgh19SN10 at $t=$0 Gyr.} \label{initMgh19}
\end{figure}

\begin{figure}
\epsscale{1.0}
\plotone{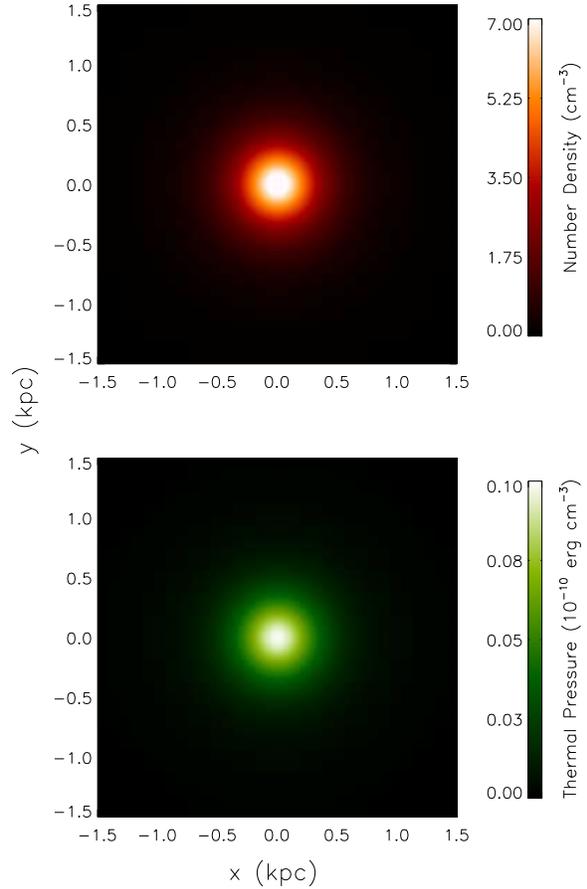}
\caption{Spatial distribution of the gas number density and pressure  (from top to bottom) on the $xy$ plane for the models Mgh05SN1 and Mgh05SN10 at $t=$0 Gyr.} \label{initMgh05}
\end{figure}

\subsubsection{The Final Setup of Our HD Models for Ursa Minor}

As mentioned above, the initial gas configuration adopted in our simulations is provided by equations (\ref{rhogas}) and (\ref{pgas}), which implies that the values of $r_0$, $\rho_\mathrm{c}$, $\rho_0$, and $c_\mathrm{s_0}$ are known a priori. However, a precise estimate of these quantities requires the knowledge of the exact properties of the gas and DM in \object{Ursa Minor} at its very early evolutionary stage, which is barely known and inferred indirectly from photometric and spectroscopic data together with stellar population and/or chemical evolution models (e.g, \citealt{kle98,kle03,mate98,bel02,car02,gre03,pal03,lama04,str07}).

We have assumed $r_0=300$ pc and $c_\mathrm{s_0} = 11.5$ km$^{-1}$ in all of the numerical simulations performed in this work. A value of 300 pc corresponds approximately to the core radius of the best-fitting King-model profile for the radial distribution of stars in \object{Ursa Minor} \citep{irha95,kle98,pal03}. A sound speed of $11.5$ km$^{-1}$ corresponds to a temperature of about 9544 K, which is lower than the threshold temperature we work out for cooling in our simulations (see the previous sections). This prevents deviations from the initial hydrostatic equilibrium condition.

According to \citet{str07}, the maximum circular velocity in \object{Ursa Minor} must be higher than 21 km s$^{-1}$. Assuming $v_\mathrm{c_\infty}=21.1$ km s$^{-1}$, we used equation (\ref{vcircinf}) to estimate $\rho_\mathrm{c}$, which led to $6.2\times 10^{-24}$ g cm$^{-3}$. The total mass of the DM halo of \object{Ursa Minor} can be obtained after integrating equation (\ref{dm_dens}) in spherical coordinates, resulting in the following:

\begin{equation} \label{dm_mass} 
M_\mathrm{h} = 4\pi\rho_\mathrm{c}r_0^3\left(\xi_\mathrm{h}-\arctan{\xi_\mathrm{h}}\right),
\end{equation}
where $\xi_\mathrm{h}=r_\mathrm{h}/r_0$ and $r_\mathrm{h}$ is the radius of the DM halo. Following \citet{mafe99}, we have assumed $r_\mathrm{h}=r_{200}$, the characteristic radius at which the mean DM density is 200 times higher than $\rho_\mathrm{crit}$, the critical density of the Universe: 

\begin{equation} \label{r_h} 
r_\mathrm{h} = \sqrt{\frac{3\rho_\mathrm{c}}{200\rho_\mathrm{crit}}},
\end{equation}
with $\rho_\mathrm{crit} = 3H_0^2/8\pi G \approx 3.7\times 10^{-29}$ g cm$^{-3}$, where $H_0$ is the Hubble constant, assumed to be equal to 71 km s$^{-1}$ Mpc$^{-1}$ in this work. Thus, we obtained $\xi_\mathrm{h}=50.0$ from equation (\ref{r_h}), leading to $M_\mathrm{h}=1.51\times 10^9$ M$_\sun$ after using equation (\ref{dm_mass}).

An estimate of the mass of the gas in \object{Ursa Minor} at initial stages, $M_\mathrm{g_0}$, is obtained from the integration of equation (\ref{rhogas}) over spatial coordinates:

\begin{equation} \label{gas_mass} 
M_\mathrm{g_0} = 4\pi\rho_0r_0^3\int\limits_{0}^{\xi_\mathrm{t}}\exp{\left[-\Gamma\frac{v^2_\mathrm{c_\infty}}{c_\mathrm{s_0}^2}\chi(\xi)\right]}d\xi,
\end{equation}
where $\xi_\mathrm{t}=r_\mathrm{t}/r_0$ and the tidal radius, $r_\mathrm{t}$, $\sim$950 pc in the case of \object{Ursa Minor} \citep{irha95}.

In principle, we can derive the value of $\rho_0$ from equation (\ref{gas_mass}) if the value of $M_\mathrm{g_0}$ is known {\it a priori}. Even though $M_\mathrm{g_0}$ is not reliably constrained by observations, some reasonable assumptions can be made concerning its value. For example, we can assume that the region where the formation of \object{Ursa Minor} took place in the past follows the primordial baryon-dark-matter ratio inferred from the fluctuations of the cosmic radiation background temperature. Using the 9 yr WMAP-only results obtained by \citet{hin13}, we have $M_\mathrm{g_0}/M_\mathrm{h}\approx 0.1956$, which implies $M_\mathrm{g_0} \approx 2.94\times 10^8$ M$_\sun$, quite similar to the initial baryonic mass assumed in the chemical evolution models for \object{Ursa Minor} \citep{lama04,lama07}. 

Inverting equation (\ref{gas_mass}) and using the values of $M_\mathrm{g_0}$, $r_0$, $\Gamma$, $v_\mathrm{c_\infty}$ and $c_\mathrm{s_0}$, we could finally obtain $\rho_0=4.6\times 10^{-23}$ g cm$^{-3}$. We have labeled our numerical simulations with $\rho_0=4.6\times 10^{-23}$ g cm$^{-3}$ and for $R_\mathrm{SNII}=1$ and 10 Myr$^{-1}$ as Mgh19SN1 and Mgh19SN10, respectively.

To check the impact of this choice on the gas loss driven by SNe II, we also performed two additional simulations considering the same SN rates of the previous models but decreasing the the value of $M_\mathrm{g_0}/M_\mathrm{h}$ from 0.1956 to 0.0526 (about a factor of 4). We labeled these as Mgh05SN1 and Mgh05SN10. A summary of the setup model parameters is provided in Table \ref{setup}. In Figures \ref{initMgh19} and \ref{initMgh05}, we present the initial spatial distribution of the gas number density and pressure on $xy$ plane for the models listed in Table \ref{setup}.

\section{RESULTS}

\subsection{General Results on Time Evolution of the Gas Content}

\begin{figure}
\epsscale{1.0}
\plotone{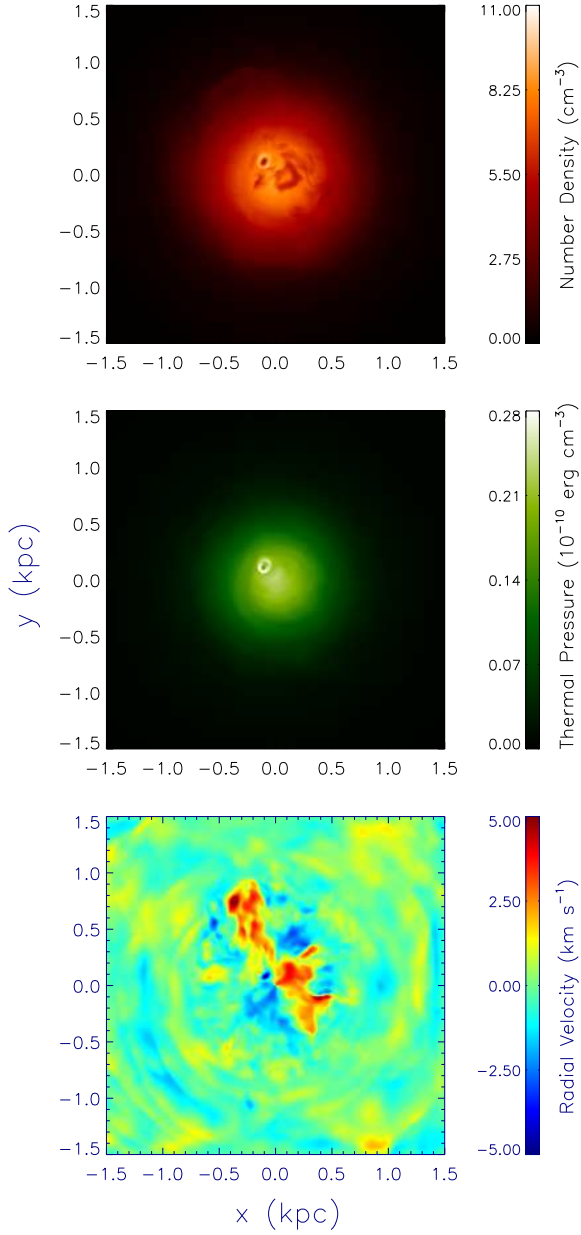}
\caption{Spatial distribution of the gas number density, thermal pressure, and radial velocity (from top to bottom) on the $xy$ plane obtained from the model Mgh19SN1 at $t=$3 Gyr. Note that the scale range of the density map was reduced by a factor of two in relation to that used in Figure \ref{initMgh19}.\label{spadistMgh19SN1}}
\end{figure}

 In Figure \ref{spadistMgh19SN1}, we show the number density, thermal pressure, and radial velocity maps of the gas on the $xy$ central slice obtained from the model Mgh19SN1 at the final snapshot $t=$3 Gyr. The value of the radial velocity, $v_\mathrm{rad}$ is calculated from

\begin{equation} \label{v_rad} 
v_\mathrm{rad} = \textbf{\emph{v}}\cdot \hat{\textbf{\emph{r}}},
\end{equation}
where $\hat{\textbf{\emph{r}}}=\textbf{\emph{r}}/\vert\textbf{\emph{r}}\vert$ is the unit position vector.

The initial isothermal spherically symmetric distribution of the gas density and pressure is strongly disturbed by the SN II explosions. The gas content is pushed outward, spreading the initial central peak seen in the density and pressure radial profiles. Irregular cavities are also created by SN blasts, as can be seen in the density plots (darker areas). We have confirmed that these low-density cavities provide low resistance and are natural and efficient channels for gas flows (e.g., \citealt{rui13}). These channels are seen in the radial velocity map of the Figure \ref{spadistMgh19SN1} as filamentary structures pointing radially to the galaxy. Outflows of gas in the central region ($v_\mathrm{rad} \ga 3$ km s$^{-1}$), as well as channels of matter flowing inward ($v_\mathrm{rad} \la -3$ km s$^{-1}$) can be noted in the same figure.

\begin{figure}
\epsscale{1.0}
\plotone{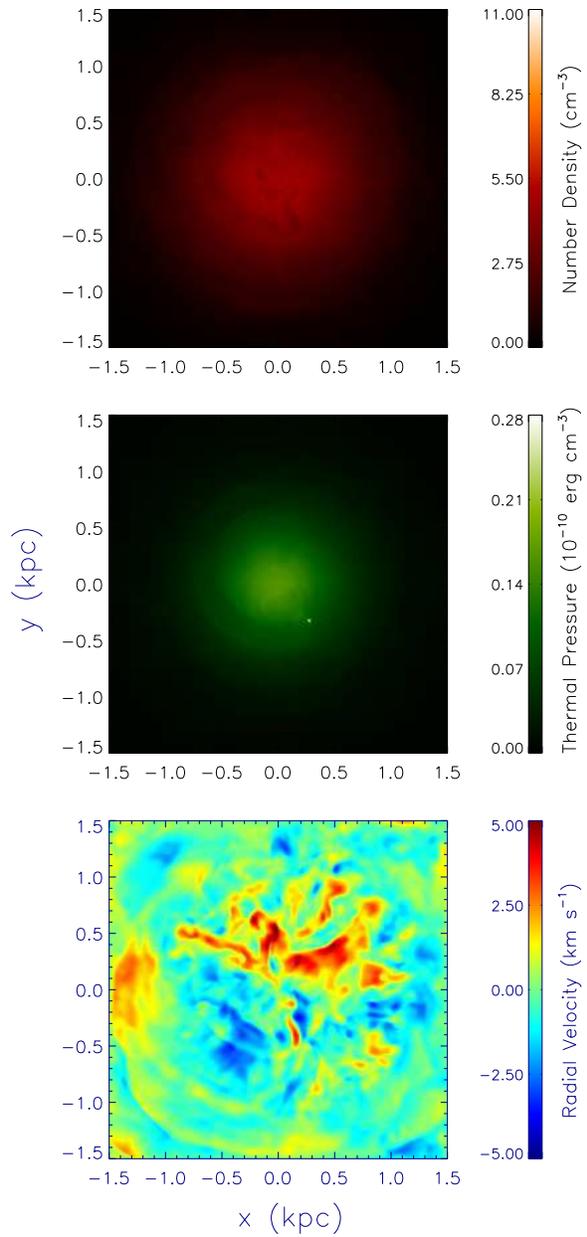}
\caption{Same as Figure \ref{spadistMgh19SN1} but for model Mgh19SN10. \label{spadistMgh19SN10}}
\vspace{2.0cm}
\end{figure}

\begin{figure}
\epsscale{1.0}
\plotone{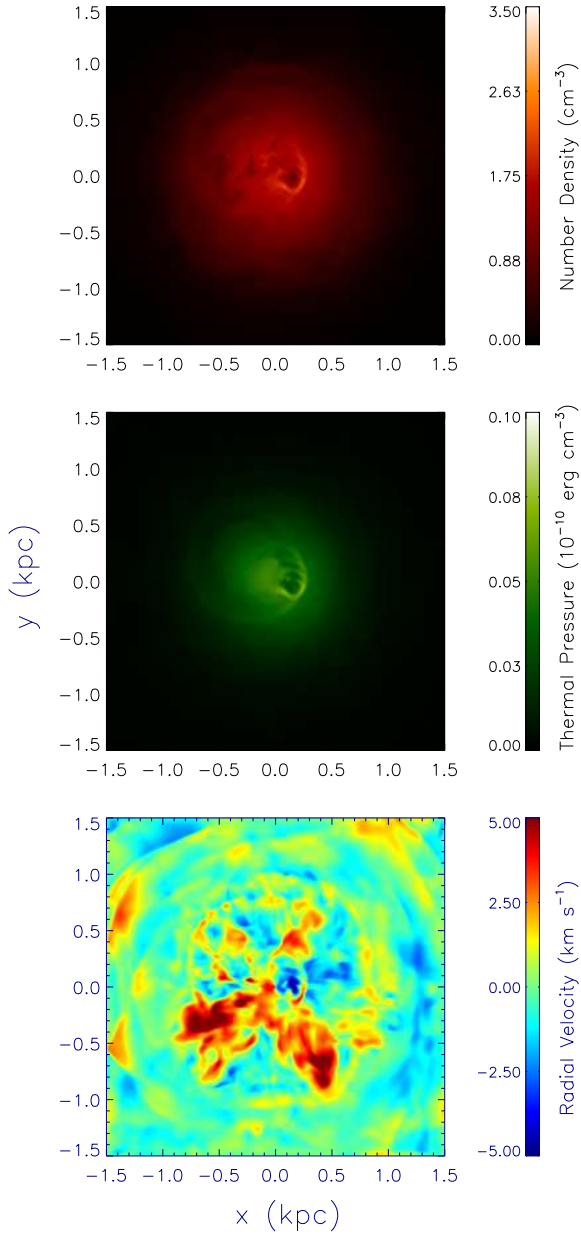}
\caption{Same as Figure \ref{spadistMgh19SN1} but for model Mgh05SN1. Note that the scale range of the density map was reduced by a factor of two in relation to that used in Figure \ref{spadistMgh19SN1}. \label{spadistMgh05SN1}}
\end{figure}

\begin{figure}
\epsscale{1.0}
\plotone{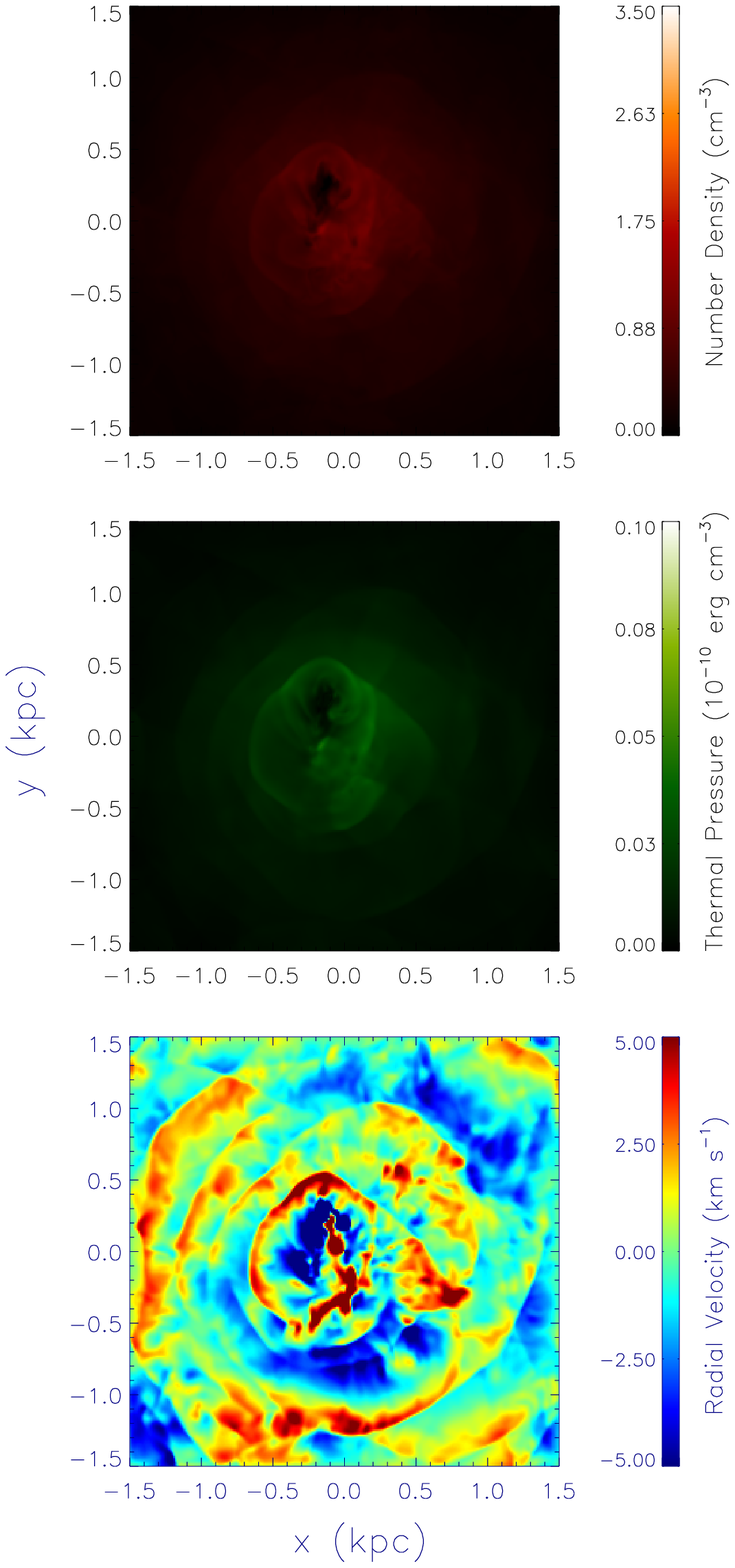}
\caption{Same as Figure \ref{spadistMgh05SN1} but for model Mgh05SN10. \label{spadistMgh05SN10}}
\vspace{1.0cm}
\end{figure}

It is important to emphasize that the resistance faced by the induced gas flows is related to the thermal pressure of the gas surrounding the SN shell. SNe shells must stall at a radius where the ram pressure is approximately equal to the local thermal pressure, which corresponds to the condition of the sonic Mach number of the shells reaching unity (e.g. \citealt{cox72,cio88,osmc88}). This happens approximately at a scale of a few tens to a hundred of parsec for the physical properties of the ISM in our model. After that, SNe remnants are expected to evolve as acoustic waves through the ISM (e.g., \citealt{spit82,fal10a,fal10b}). Indeed, there is a complex distribution of shells with radii of a few hundreds of parsecs and subsonic velocities (roughly between 1 and 2 km s$^{-1}$) seen in the radial velocity map of the Figure \ref{spadistMgh19SN1}, indicating the presence of such acoustic waves in the ISM of \object{Ursa Minor}. 

Those shell-like patterns are also present in the pressure distribution displayed in Figure \ref{spadistMgh19SN1}. The small ring-like structure seen in Figure \ref{spadistMgh19SN1} (mainly in pressure) was produced by an individual SN event not so far from the $xy$ plane (more precisely, $\sim$ 18 pc below it). This feature has a density contrast of about 1.5, a signature of a weak shock front.

Figures \ref{spadistMgh19SN10}, \ref{spadistMgh05SN1} and \ref{spadistMgh05SN10} are the same as Figure \ref{spadistMgh19SN1} but referring, respectively, to the models Mgh19SN10, Mgh05SN1, and Mgh05SN10. The features seen in the maps of Figure \ref{spadistMgh19SN1} are also found in those figures, even though they are not necessarily identical. For instance, the gas content inside a radius of 500 pc becomes sparser and less peaked after 3 Gyr when the SN rate is increase (e.g., compare the extension of the spatial distribution of the gas with $n\ga 3$ cm$^{-3}$ in the Figures \ref{spadistMgh19SN1} and \ref{spadistMgh19SN10}). This was already expected because more thermal energy is injected into the gas, increasing also the kinetic energy available to push it outward (e.g., \citealt{mafe99}). The same trend is observed when the initial gas density is lowered (e.g., compare the extension of the spatial distribution of the gas with $n\ga 3$ cm$^{-3}$ in the Figures \ref{spadistMgh19SN10} and \ref{spadistMgh05SN10}). This result is also expected since low-density environments offer less resistance to the gas motions driven by the SN blasts, which is also quantitatively corroborated by the systematic higher absolute values of $v_\mathrm{rad}$ shown in Figure \ref{spadistMgh05SN10}. As it will be discussed in the next section, both results play an important role in the mass-loss process in \object{Ursa Minor}.

The radial velocity maps of the gas shown in Figures \ref{spadistMgh19SN10} -- \ref{spadistMgh05SN10} display the same features seen in the $v_\mathrm{rad}$-map of Figure \ref{spadistMgh19SN1}.
Note that the larger changes in the values of $v_\mathrm{rad}$ occur in the vicinity of low-density cavities and/or shock fronts seen in the density/pressure plots. In addition, regions with high radial velocities (i.e., $\vert v_\mathrm{rad}\vert\ga 5$ km s$^{-1}$) are more abundant for the highest SN rate assumed in this work ($R_\mathrm{SNII}=10$ Myr$^{-1}$; Figures \ref{spadistMgh19SN10} and \ref{spadistMgh05SN10}), as well as when the initial average gas density is lower (e.g., compare Figures \ref{spadistMgh19SN1} and \ref{spadistMgh05SN1}).

In summary, we have noted two different patterns related to the radial velocities observed in our four numerical simulations: a filamentary structure with velocities higher than $\sim 3$ km s$^{-1}$, and the superposition of concentric shells with velocities in the range of about $1-3$ km s$^{-1}$. A possible explanation for the filamentary distribution of $v_\mathrm{rad}$ is related to the Rayleigh-–Taylor instability (RTI). In this process, SN explosions generate cavities of hot and diffuse gas which then suffer from the buoyancy effect. As the cavities/bubbles rise upward from the central region of the galaxy, they become stretched in filamentary morphologies, similar to the "mushroom-like" structures caused by nuclear explosions. One can estimate the radial uprise velocity of such structures as (see \citealt{rui13} for details)

\begin{equation} \label{v_RT} 
v_\mathrm{RT}(t) \simeq 0.1 gt \left(\frac{\rho_\mathrm{ISM}-\rho_\mathrm{cav}}{\rho_\mathrm{ISM}+\rho_\mathrm{cav}}\right),
\end{equation}
where $\rho_\mathrm{ISM}$ and $\rho_\mathrm{cav}$ represent the gas densities of the surrounding ISM and the cavities, respectively, and $g$ the local acceleration of gravity from the DM and baryonic mass components. For the parameters obtained in our models, we find, at typical times of $\sim$ 100 Myr, $v_\mathrm{RT} \sim 4-5$ km s$^{-1}$, in absolute agreement with the velocities observed in the filaments. Therefore, it is probable that RTI is the dominant process generating these structures.

\begin{figure}
\epsscale{0.91}
\plotone{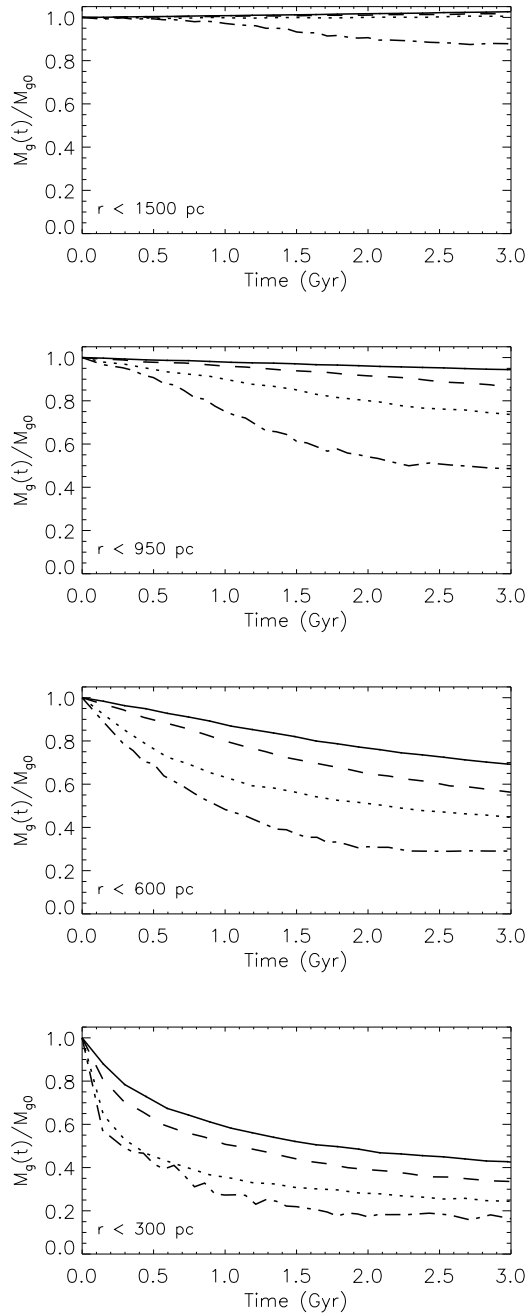}
\caption{From top to bottom, we show the instantaneous mass of the gas normalized by that at time zero inside a spherical radius smaller than 1.5, 0.9, 0.6, and 0.3 kpc, respectively. Solid lines represent the results from model Mgh19SN1, while the dotted, dashed, and dashed-dotted lines refer, respectively, to models Mgh19SN10, Mgh05SN1, and Mgh05SN10. \label{massloss}}
\end{figure}

Concerning the concentric shells, they have velocities smaller than the local sound speed, i.e. at Mach numbers of $M_\mathrm{s} \equiv \langle\delta v\rangle/c_\mathrm{s} \sim 0.1 - 0.3$ (e.g., see Figure \ref{spadistMgh05SN10}). Such behavior suggests that the shells are related to acoustic waves (e.g., \citealt{spit82}), as \citet{fal10b} had already been pointed out in the context of massive galaxies in clusters. As the sound waves propagate radially, their amplitudes decay as does the wave action, resulting in a net force outward. The semi-analytical solution for a wave-driven galactic wind, assuming an isothermal and isotropic wind with similar parameters found in model Mgh05SN10, result in a specific mass-loss rate, $\dot{M}/n(r=200$ $\mathrm{pc})$, equal to $\sim 10^{-3}$ M$_\odot$ yr$^{-1}$ cm$^3$ (see \citealt{falc13}). Therefore, theoretically, on a timescale of 1 Gyr, a total mass of $\sim 5\times10^6$ M$_\odot$ would have been pushed outward by the wave pressure in model Mgh05SN10, which corresponds to $\sim 10$\% of the total mass internal to $r=900$ pc. This value is half of the mass loss observed within this radius in this simulation after 1 Gyr (see Figure \ref{massloss} in Section 3.2). Combined with the RTI, both processes explain quantitatively well the mass loss inferred from this numerical simulation.

\subsection{Mass-loss Efficiency and SNe Rates}

In Figure \ref{massloss}, we show the time evolution of the total barionic mass of the gas in the ISM of \object{Ursa Minor}, normalized by its initial value. The relative masses were obtained by  integrating the total gas mass within galactocentric radii of 0.3, 0.6, 0.95, and 1.5 kpc. 

For a spherical region with a radius of 1.5 kpc, which corresponds to almost the entire computational domain, Mgh05SN10 is the only model presenting a considerably large mass loss ($\sim$12\%) after 3 Gyr. The other models present roughly no mass loss (model Mgh05SN1) or some accretion ($<3$\% in the worst case). We attribute the inferred increase of the enclosed mass in models Mgh19SN1 and Mgh19SN10 to a combination of two effects. 

\begin{enumerate}

\item[i.] The inefficiency of blowing gas out due to the intrinsic higher gas densities in these models. Note that SN-driven perturbations advance to outer regions more efficiently in the case of the lower-density models (Mgh05SN1 and Mgh05SN10), as shown, for instance, in the Figures \ref{spadistMgh19SN1} -- \ref{spadistMgh05SN10}.

\item[ii.] Deviations from the perfect hydrostatic equilibrium between gas and the DM gravitational field due to domain discretization (see section 2.2). Note that spurious mass increments have always been smaller than the value (6.1\%) found in our stability test regarding hydrostatic equilibrium discussed in the section 2.2.2.

\end{enumerate}

All models present gas loss for smaller radii (0.3, 0.6, and 0.95 kpc). It is evident that the higher the SN rate, the higher the mass loss. The same trend is observed for the gas density: lower-density models are prone to lose larger amounts of gas, as seen in previous works for generic dwarf galaxies (e.g. \citealt{buru97,mafe99,feto00,fra04}). However, besides considering a specific dSph galaxy and letting the SNe explosions distribute over space and time (as in \citealt{rui13}), we have also explored in this work how the relative mass loss is dependent on the radius defined to integrate it. The larger density of the SNe events in the the central regions of the galaxy results in the radial dependence of the mass loss. We also observed in our simulations that the gas is not completely removed, but pushed to for \object{Ursa Minor}'s envelope. Maximum mass losses of 65-85\% of the initial total mass were obtained inside 300 pc. For a volume delimited by a radius of 600 pc, where the majority of stars in \object{Ursa Minor} has been detected (e.g., \citealt{irha95}), the relative mass loss ranges from 30 to 70\%. The gas loss decreases to about 5--50\% inside 950 pc, the approximate tidal radius of \object{Ursa Minor} (e.g., \citealt{irha95}).

As was already discussed in the previous section, the combinations of the RTI and the SN-driven acoustic waves can account for the mass loss inferred from our numerical simulations.

\subsection{Comparison between Numerical Simulations and Observational Constraints}

Our HD simulations of the gas content in \object{Ursa Minor} were performed assuming a constant SN rate during the whole 3 Gyr. As mentioned in Section 2, those rates generated a total number of SNe at the end of the simulations compatible with those expected from chemical evolution models for \object{Ursa Minor} \citep{lama04}, as well as a star formation history compatible with the inferred value of about 8$\times 10^5$ M$_\sun$ in stars for \object{Ursa Minor} \citep{dewo03,orb08}.

Our numerical simulations showed that the adopted SN II rates (1 and 10 Myr$^{-1}$) were able to transfer most of the gas from the central region ($<300$ pc) outward to the galactic halo. However, SN II feedback did not completely remove the gas from \object{Ursa Minor}. A total mass of $1.4\times 10^7$ M$_\sun$ remained inside a radius of 600 pc after 3 Gyr. This is two orders of magnitude higher than the upper limit of $\sim 10^5$ M$_\sun$ for \ion{H}{2} mass at the present time derived by \citet{gal03} from H$\alpha$ observations. Tentative detections of the 21 cm \ion{H}{1} emission line in \object{Ursa Minor} put an upper limit of about 7000 M$_\sun$ for the amount of neutral hydrogen \citep{youn00,gre03}, again too small compared to the remaining mass found from our simulations.

This mass excess can also be noted in terms of the column density calculated from the spatial integration of the gas density along the $z$ direction. After 3 Gyr, column densities inside a radius of 600 pc from the galactic nucleus are higher than $10^{22}$ cm$^{-2}$ in simulations Mgh19SN1 and Mgh19SN10, decreasing to $\sim 6\times 10^{21}$ cm$^{-2}$ and $\sim 3\times 10^{21}$ cm$^{-2}$ in the cases of simulations Mgh05SN1 and Mgh05SN10, respectively. These values are higher than the upper limit of about $5\times 10^{17}$ cm$^{-2}$ for the \ion{H}{1} column density derived by \citet{youn00}.

Therefore, our results suggest that SN II feedback alone is insufficient to completely remove the ISM of \object{Ursa Minor} under the physical conditions adopted in this work, even though galactic winds have blown out a substantial amount of gas in our simulations (between 30 and 70\% inside 600 pc). To reconcile or at least minimize the differences between our numerical results and observational constraints on the gas content in \object{Ursa Minor}, some additional mechanism and/or different initial gas/DM conditions are necessary. We list some possible and non-exclusive candidates in the next sections.

\subsubsection{Gas-to-DM Ratio}

Simulations Mgh05SN1 and Mgh05SN10 started with $\sim 8\times 10^7$ M$_\sun$ in gas inside 950 pc, a factor of $\sim 3.7$ lower than the initial gas mass in the other two simulations. It increased the mass-loss efficiency within 600 pc in $\sim$23\% for $R_\mathrm{SNII}=1$ Myr$^{-1}$, and $\sim$52\% for $R_\mathrm{SNII}=10$ Myr$^{-1}$. These results suggest decreasing the gas-to-DM ratio as a possible way to bring the final gas mass closer to the observational limits for \object{Ursa Minor}. The main consequence of this scenario is that \object{Ursa Minor} could have been born in a relatively low-density environment in the Local Group. 

An additional possibility could be a smaller initial mass of the DM halo of \object{Ursa Minor} with a time-dependent growth (from hierarchical accretion and merging of DM sub-halos) until reaching its present-day inferred value. Such behavior is expected from theoretical and numerical works in the context of $\Lambda$CDM\footnote{Acronym of $\Lambda$ Cold Dark Matter.} cosmological models (e.g., \citealt{whre78,dav85,kly99,vog14}). Indeed, \citet{saw10} have shown (from SPH numerical simulations) that the time-dependent growth of a DM halo together with SN feedback and an external UV radiation field can produce galaxies with structural parameters similar to dwarf galaxies in the Local Group.

\subsubsection{SNe Ia}

We have ignored the feedback from SNe Ia in our simulations since they are usually less frequent than SNe II (e.g., \citealt{lama04,mar06}). However, their contribution to the mass-loss process may not be negligible.

SNe Ia occur long (1 Gyr or so) after the evolution of low-mass stars in binary systems, even during the quiescent phases of SNe II, injecting energy into the gas and probably contributing to the gas loss even after the time our simulations are halted. In fact, in the chemical evolution models by \citet{lama04,lama07}, the SNe Ia explosions still occur at very recent epochs, but at a very low rate, in \object{Ursa Minor}. They could also maintain \object{Ursa Minor}'s ISM at higher temperatures, allowing it to be less bound to the DM gravitational potential well (e.g., \citealt{mar06,saw10}).

\subsubsection{Lower-density Intergalactic Medium (IGM)}

To avoid introducing a strong discontinuity to the gas density distribution in our simulations, we have used equation \ref{rhogas} beyond the tidal radius of \object{Ursa Minor}. It produced an external environment with a density $\sim 10^{4-5}$ higher than typical estimates for the IGM across the Local Group at present \citep{mura00,grpu09,gat13}. These denser environments decelerate the galactic winds driven by the SNe blasts more efficiently and, as a consequence, lower the mass-loss efficiency. Note that this over-dense IGM also impacts the estimates of column densities from our simulations.

\subsubsection{UV Background Radiation Field}

Some previous works have addressed the importance of the UV background radiation field to the gas-loss process in the context of the dwarf galaxies (e.g., \citealt{rea06,saw10,sim13,mibr14}). For example, \citet{saw10} and \citet{sim13} showed that the UV radiation field is primarily responsible for expelling most of the low-density intergalactic gas, even though the dense, cold gas in the core of the halo is mainly dispersed by SN feedback.

The contribution of the UV radiation field in our simulations is taken into account in the radiative cooling function, as well as in its suppression below a threshold temperature (see section 2.2). However, the dependence on UV heating due to an external background has not been studied in detail as a free parameter in our calculations, which could eventually increase the mass-loss rates (e.g., \citealt{kat96,hoe06,saw10,sim13,falc13}).

\subsubsection{Tidal Stripping}

In general, tidal effects become important when the gravitational binding acceleration of a satellite galaxy is similar to the differential acceleration exerted on it due to the host galaxy (e.g., \citealt{blro00,rea06a,rea06b}). This leads to the concept of the tidal radius (defined in section 2.2), a critical radius beyond which matter can be tidally stripped from the satellite by the host galaxy.

\citet{rea06a} showed through analytical calculations and $N$-body simulations that $r_\mathrm{t}$ depends on the gravitational potentials of the host and satellite galaxies, as well as on the orbit of the satellite around the host and the orbit of the stars within the satellite. \citet{gre03}, \citet{rea06b} and \citet{gat13} argued that tidal stripping is unimportant inside a galactic radius of about 1 kpc for most of the Local Group dSph galaxies observed up to now. However, these calculations are based only on dynamical considerations, ruling out any non-gravitational energy release inside the tidal radius influence. For example, the energy released by SNe can weaken the gravitational potential, changing the DM density profiles from cusp to core during the galaxy's evolution (e.g, \citealt{mas06,pas10,pon12}).

The effects of the tidal stripping mechanism on the gas removal in \object{Ursa Minor} will be investigated in future works.

\subsubsection{Ram-pressure Stripping}

Galaxies moving through the IGM are subject to the ram-pressure stripping if \citep{gugo72,gre03}:

\begin{equation} \label{rampressure} 
\rho_\mathrm{IGM} v_\mathrm{gal}^2 \ga \rho_\mathrm{gal}\sigma_\mathrm{gal}^2,
\end{equation}
where $\rho_\mathrm{IGM}$ is the intergalactic gas density, $\sigma_\mathrm{gal}$ is the galactic velocity dispersion, and $\rho_\mathrm{gal}$ and $v_\mathrm{gal}$ are, respectively, the mean density and the velocity of the galaxy in relation to IGM. This simple analytical expression is supported by more detailed HD simulations (e.g. \citealt{mcc07,may06}; see \citealt{gat13} for some criticism), even though it is formally valid only for a dimensionless galaxy (see \citealt{pas12,pas15} for a generalization of equation \ref{rampressure} in the case of a galaxy with a non-zero size).

In the case of \object{Ursa Minor}, the mean gas number density (mean column density inside 600 pc) must be smaller than about $2\times10^{-2}$ cm$^{-3}$ ($7\times10^{19}$ cm$^{-2}$) in order for ram-pressure stripping to remove efficiently its gas component. We have adopted in these conservative calculations $\rho_\mathrm{IGM}\approx 1.3\times10^{-4}$ cm$^{-3}$, corresponding to the minimum average particle density expected at distances smaller than 90 kpc from the Milky Way \citep{grpu09,gat13}, $v_\mathrm{gal}\approx 162$ km s$^{-1}$ \citep{pia05}, and $\sigma_\mathrm{gal} \approx 12$ km s$^{-1}$ \citep{wil04}.

We can realize that the derived upper limit of $2\times10^{-2}$ cm$^{-3}$ is at least a factor of 26 smaller than the mean number density found in our numerical simulations after 3 Gyr of evolution. At a first glance, this indicates that ram-pressure stripping could play a small role in removing gas from \object{Ursa Minor}. However, Equation (\ref{rampressure}) does not take into account nonlinear effects behind gas-shocking interactions, which could change our naive estimate for that critical density \citep{gat13}. Besides, a decrease in the initial gas-to-DM ratio could increase the amount of gas stripped by ram-pressure effects.

We will address the effect of the ram-pressure stripping mechanism in the context of \object{Ursa Minor} in a forthcoming work.

\section{CONCLUSIONS}

In this work, we presented the results from 3D HD simulations of the gas content of the dwarf galaxy \object{Ursa Minor}, emulating 3 Gyr of its evolution. So far, this is the first direct attempt to model the gas mass evolution of a particular dSph galaxy of the Local Group in terms of 3D HD simulations. We used a computational grid of 256$^3$ points, distributed uniformly across a box of 3$^3$ kpc$^3$, resulting in a spatial resolution of about 11.7 pc.

Initially, the isothermal gas is placed in hydrostatic equilibrium with a cored, static DM gravitational field. The total DM mass was constrained by the minimum value of the total velocity dispersion of \object{Ursa Minor}. We also assumed two different values for the gas-to-DM mass ratio: about 0.20 and 0.05, the former compatible with the 9 yr WMAP-only results \citep{hin13}. These initial equilibrium configurations were disturbed by SN II explosions, at constant rates of 1 and 10 Myr$^{-1}$, in agreement with the estimates from chemical evolution models for this object \citep{lama04,lama07}.

From the star formation history for the dSph galaxy \object{Ursa Minor} assumed in this work, we conclude the following.

\begin{enumerate}

\item[i.] Gas is spread outward, erasing the initial central peak in the density and pressure distributions. Irregular, low-density cavities were also created by SN-driven shocks. The highest spatially averaged radial velocities were found in the case of the highest SN rate assumed in this work.

\item[ii.] Filamentary structures seen in the radial velocity maps ($v_\mathrm{rad} \ga 3$ km s$^{-1}$) and possibly induced by the RTI are predominantly associated with gas outflows. Some channels of matter flowing inward ($v_\mathrm{rad} \la -3$ km s$^{-1}$) are also seen in our simulations, as well as a complex distribution of shell-like patterns with subsonic velocities roughly between 1 and 2 km s$^{-1}$, associated with the propagation of acoustic waves through the ISM.

\item[iii.] The efficiency of the gas removal is higher when the SN rate is increased, as expected because more thermal energy is injected into the gas. The same trend is observed when the initial gas density is lowered, since low-density environments offer less resistance to the gas motions and the energy is redistributed over less mass, resulting in larger outward velocities.

\item[iv.] The induced gas loss by SNe is differential in terms of the galactic radius and variable in time, independent of the model parameters considered in this work (see Figure \ref{massloss}). After 3 Gyr, the derived mass-loss rates reached their maximum inside 300 pc, with values ranging from about 65 to 85\% of the initial mass depending on the assumed initial gas density and SN II rate. For a spherical radius smaller than 600 pc, where the majority of stars in \object{Ursa Minor} has been detected (e.g., \citealt{irha95}), the relative mass-loss rates ranged from 30 to 70\%. The gas loss decreases to about 5\%--50\% inside 950 pc, roughly the tidal radius of \object{Ursa Minor}.

\item[v.] We show from semi-analytical calculations that the combination of RTI and acoustic waves driven by the SNe can explain quantitatively the mass loss in \object{Ursa Minor} inferred from our simulations.

\item[vi.] Even though galactic winds have blown out substantial amount of gas in our simulations (between 30 and 70\% inside 600 pc), our results suggest that SN II feedback alone was insufficient to remove completely the ISM of \object{Ursa Minor}.

\end{enumerate}

In order to completely remove the gas at larger radii, other additional internal and/or external mechanisms must be considered. Possible and non-exclusive candidates may be a different initial gas-to-DM ratio, a lower-density IGM surrounding \object{Ursa Minor}, as well as the inclusion of a UV background radiation field, ram-pressure and tidal stripping effects, and SN I feedback. We plan to study the influence of these additional mechanisms in the context of \object{Ursa Minor} galaxy in a forthcoming work.



\acknowledgments

This work has made use of the computing facilities of the Laboratory of Astroinformatics (IAG/USP, NAT/UCS), whose purchase was made possible by the Brazilian agency FAPESP (grant 2009/54006-4) and the INCT-A. A.C. thanks the S\~ao Paulo Research Foundation (FAPESP) for financial support (grant \#2015/06361-0). G.A.L. thanks CNPq (grant \#308677/2012-9) for financial support. D.F.G. thanks the European Research Council (ADG-2011 ECOGAL), and Brazilian agencies CAPES (3400-13-1) and FAPESP (grant \#2011/12909-8) for financial support. The authors also thank the anonymous referee for a detailed and careful report that improved the presentation of this work.

\clearpage

\clearpage





\begin{thebibliography}{}

\bibitem[Battaglia et al.(2006)]{bat06} Battaglia, G., Tolstoy, E., Helmi, A. et al. 2006, \aap, 459, 423

\bibitem[Bellazzini et al.(2002)]{bel02} Bellazzini, M., Ferraro, F. R., Origlia, L., et al. 2002, \aj, 124, 3222

\bibitem[Binney \& Tremaine(1987)]{bitr87} Binney, J., \& Tremaine, S. 1987, Galactic Dynamics (Princeton, NJ: Princeton Univ. Press), 601

\bibitem[Blitz \& Robishaw(2000)]{blro00} Blitz, L., \& Robishaw, T. 2000, \apj, 541, 675

\bibitem[Bosch-Ramon et al.(2012)]{bos12} Bosch-Ramon, V., Barkov, M. V., Khangulyan, D., \& Perucho, M. 2012, \aap, 544, 59

\bibitem[Burkert(1995)]{burk95} Burkert, A. 1995, ApJL, 543, L23

\bibitem[Burkert \& Ruiz-Lapuente(1997)]{buru97} Burkert, A., \& Ruiz-Lapuente, P. 1997, \apj, 480, 297

\bibitem[Burkert et al.(2012)]{bur12} Burkert, A., Schartmann, M., Alig, C., et al. 2012, \apj, 750, 58

\bibitem[Carrera et al.(2002)]{car02} 
Carrera, R., Aparicio, A., Martínez-Delgado, D., \& Alonso-García, J. 2002, \aj, 123, 3199

\bibitem[Cioffi, McKee, \& Bertschinger(1988)]{cio88} Cioffi, D. F., McKee, C. F., Bertschinger, E. 1988, \apj, 334, 252

\bibitem[Cox(1972)]{cox72} Cox, D. P. 1972, \apj, 178, 159

\bibitem[Davis et al.(1985)]{dav85} Davis, M., Efstathiou, G., Frenk, C. S., \& White, S. D. M. 1985, \apj, 292, 371

\bibitem[de Blok \& Bosma(2002)]{blbo02} de Blok, W. J. G., \& Bosma, A. 2002, \aap, 385, 816

\bibitem[de Boer et al.(2012a)]{boe12a} de Boer, T. J. L., Tolstoy, E., Hill, V., et al. 2012a, \aap, 539, 103

\bibitem[de Boer et al.(2012b)]{boe12b} de Boer, T. J. L., Tolstoy, E., Hill, V., et al. 2012b, \aap, 544, 73

\bibitem[Dekel \& Woo(2003)]{dewo03} Dekel, A., \& Woo, J. 2003, \mnras, 344, 1131

\bibitem[Del Popolo(2012)]{popo12} Del Popolo, A. 2012, \mnras, 419, 971

\bibitem[Dolphin et al.(2005)]{dol05} Dolphin, A. E., Weisz, D. R., Skillman, E. D., \& Holtzman, J. A. 2005, arXiv:astro-ph/0506430

\bibitem[Falceta-Gon\c{c}alves et al.(2010a)]{fal10a} Falceta-Gon\c{c}alves D., Caproni A., Abraham Z., Teixeira D. M., de Gouveia Dal Pino E. M., 2010a, ApJL, 713, L74

\bibitem[Falceta-Gon\c{c}alves et al.(2010b)]{fal10b} Falceta-Gon\c {c}alves, D., de Gouveia Dal Pino, E. M., Gallagher, J. S., 
\& Lazarian, A.\ 2010b, ApJL, 708, L57

\bibitem[Falceta-Gon\c{c}alves(2013)]{falc13} Falceta-Gon\c{c}alves, D. 2013, \mnras, 432, 589

\bibitem[Ferrara \& Tolstoy(2000)]{feto00} Ferrara, A., \& Tolstoy, E. 2000, \mnras, 313, 291

\bibitem[Fragile et al.(2003)]{fra03} Fragile, P. C., Murray, S. D., Anninos, P., \& Lin, D. N. C. 2003, \apj, 590, 778

\bibitem[Fragile, Murray \& Lin(2004)]{fra04} Fragile, P. C., Murray, S. D., \& Lin, D. N. C. 2004, \apj, 617, 1077

\bibitem[Gallagher et al.(2003)]{gal03} Gallagher, J. S., Madsen, G. J., Reynolds, R. J., Grebel, E. K., Smecker-Hane, T. A. 2003, \apj, 588, 326

\bibitem[Gatto et al.(2013)]{gat13} Gatto, A., Fraternali, F., Read, J. I. et al. 2013, \mnras, 433, 2749

\bibitem[Governato et al.(2010)]{gov10} Governato, F., Brook, C., Mayer, L., et al. 2010, \nat, 463, 203

\bibitem[Grebel, Gallagher \& Harbeck(2003)]{gre03} Grebel, E. K., Gallagher, J. S., \& Harbeck, D. 2003, \aj, 125, 1926

\bibitem[Grebel(2008)]{gre08} Grebel, E. K. 2008, in IAU Symp. 244, Dark Galaxies and Lost Baryons, ed. J. I. Davies, \& M. J. Disney (Cambridge: CUP), 300 

\bibitem[Grcevich \& Putman(2009)]{grpu09} Grcevich J., \& Putman M. E., 2009, \apj, 696, 385

\bibitem[Gunn \& Gott(1972)]{gugo72} Gunn, J. E., \& Gott, J. R. III 1972, \apj, 176, 1

\bibitem[Haardt \& Madau(2001)]{hama01} Haardt, F., \& Madau, P. 2001, in Clusters of Galaxies and the High Redshift Universe Observed in X-Rays, ed. D. M. Neumann \& J. T. T. Van (CEA Saclay), 64

\bibitem[Hinshaw et al.(2013)]{hin13} Hinshaw, G., Larson, D., Komatsu, E., et al. 2013, \apjs, 208, 19

\bibitem[Hoeft et al.(2006)]{hoe06} Hoeft M., Yepes G., Gottl\" ober S., \& Springel V. 2006, \mnras, 371, 401

\bibitem[Irwin \& Hatzidimitriou(1995)]{irha95} Irwin, M., \& Hatzidimitriou, D. 1995, \mnras, 277, 1354

\bibitem[Jardel \& Gebhardt(2012)]{jage12} Jardel, J. R., \& Gebhardt, K. 2012, \apj, 746, 89

\bibitem[Katz(1992)]{katz92} Katz, N. 1992, \apj, 391, 502

\bibitem[Katz, Weinberg, \& Hernquist(1996)]{kat96} Katz, N., Weinberg, D. H., \& Hernquist, L. 1996, \apjs, 105, 19

\bibitem[Kay et al.(2002)]{kay02} Kay, S. T., Pearce, F. R., Frenk, C. S., \& Jenkins, A. 2002, \mnras, 330, 113

\bibitem[Kennicutt(1998)]{kenn98} Kennicutt R. C. 1998, \apj, 498, 541

\bibitem[Kirby et al.(2011)]{kir11} Kirby, E. N., Lanfranchi, G. A., Simon, J. D., Cohen, J. G., \& Guhathakurta, P. 2011, \apj, 727, 78

\bibitem[Kleyna et al.(1998)]{kle98} 
Kleyna, J. T., Geller, M. J., Kenyon, S. J., Kurtz, M. J., \& Thorstensen, J. R. 1998, \aj, 115, 2359

\bibitem[Kleyna et al.(2003)]{kle03} Kleyna, J. T., Wilkinson, M. I., Gilmore, G., \& Evans, N. W. 2003, ApJL, 588, L21

\bibitem[Klypin et al.(1999)]{kly99} Klypin, A., Kravtsov, A. V., Valenzuela, O., \& Prada, F. 1999, \apj, 522, 82

\bibitem[Lanfranchi \& Matteucci(2003)]{lama03} Lanfranchi, G. A., \& Matteucci, F. 2003, \mnras, 345, 71

\bibitem[Lanfranchi \& Matteucci(2004)]{lama04} Lanfranchi, G. A., \& Matteucci, F. 2004, \mnras, 351, 1338

\bibitem[Lanfranchi \& Matteucci(2007)]{lama07} Lanfranchi, G. A., \& Matteucci, F. 2007, \aap, 468, 927

\bibitem[Mac Low \& Ferrara(1999)]{mafe99} Mac Low, M-M., \& Ferrara, A. 1999, \apj, 513, 142

\bibitem[Marcolini et al.(2006)]{mar06} Marcolini, A., D'Ercole, A., Brighenti, F., Recchi, S. 2006, \mnras, 371, 643

\bibitem[Mashchenko, Couchman \& Wadsley(2006)]{mas06} Mashchenko, S., Couchman, H. M. P., Wadsley, J. 2006, \nat, 442, 539

\bibitem[Mateo(1998)]{mate98} Mateo, M. L. 1998, \araa, 36, 435

\bibitem[Mayer et al.(2006)]{may06} Mayer L., Mastropietro C., Wadsley J., Stadel J., \& Moore B. 2006, \mnras, 369, 1021

\bibitem[Mayer et al.(2007)]{may07} Mayer, L., Kazantzidis, S., Mastropietro, C., \& Wadsley, J. 2007, \nat, 445, 738

\bibitem[McConnachie et al.(2007)]{mcc07} McConnachie A. W., Venn K. A., Irwin M. J., Young L. M., \& Geehan J. J. 2007, ApJL, 671, L33

\bibitem[Mignone et al.(2007)]{mig07} Mignone, A., Bodo, G., Massaglia, S., et al. 2007, \apjs, 170, 228

\bibitem[Mignone et al.(2012)]{mig12} Mignone, A., Flock, M., Stute, M., Kolb, S. M., \& Muscianisi, G. 2012, \aap, 545, 152

\bibitem[Milosavljevi\' c \& Bromm(2014)]{mibr14} Milosavljevi\' c, M., \& Bromm, V. 2014, \mnras, 440, 50

\bibitem[Mori \& Umemura(2006)]{moum06} Mori, M., \& Umemura, M. 2006, \nat, 440, 644

\bibitem[Murali(2000)]{mura00} Murali, C. 2000, ApJL, 529, L81

\bibitem[Oh et al.(2011)]{oh11} Oh S., Brook C., Governato, F., et al. 2011, \aj, 142, 24

\bibitem[Orban et al.(2008)]{orb08} Orban, C., Gnedin, O. Y., Weisz, D. R., et al. 2008, \apj, 686, 1030

\bibitem[Ostriker \& McKee(1988)]{osmc88} Ostriker, J. P., \& McKee, C. F. 1988, Rev. Mod. Phys., 60, 1

\bibitem[Palma et al.(2003)]{pal03} Palma, C., Majewski, S. R., Siegel, M. H., et al. 2003, \aj, 125, 1352

\bibitem[Pasetto et al.(2010)]{pas10} Pasetto, S., Grebel, E. K., Berczik, P., Spurzem, R., Dehnen, W. 2010, \aap, 514, 47

\bibitem[Pasetto et al.(2012)]{pas12} Pasetto, S., Bertelli, G., Grebel, E. K., Chiosi, C., Fujita, Y. 2012, \aap, 542, 17

\bibitem[Pasetto et al.(2015)]{pas15} Pasetto, S., Cropper, M., Fujita, Y., Chiosi, C., Grebel, E. K. 2015, \aap, 573, 48

\bibitem[Piatek et al.(2005)]{pia05} Piatek, S., Pryor, C., Bristow, P., et al. 2005, \aj, 130, 95

\bibitem[Pontzen \& Governato(2012)]{pon12} Pontzen, A., Governato, F. 2012, \mnras, 421, 3464

\bibitem[Porth \& Fendt(2010)]{pofe10} Porth, O., \& Fendt, C. 2010, \apj, 709, 1100

\bibitem[Qian \& Wasserburg(2012)]{qiwa12} Qian, Y.-Z., \& Wasserburg, G. J. 2012, Proceedings of the National Academy of Sciences, 109, 4750

\bibitem[Read, Pontzen \& Viel(2006)]{rea06} Read J. I., Pontzen A. P., \& Viel M. 2006, \mnras, 371, 885

\bibitem[Read et al.(2006a)]{rea06a} Read, J. I., Wilkinson, M. I., Evans, N. W., Gilmore, G., \& Kleyna, J. T. 2006a, \mnras, 366, 429

\bibitem[Read et al.(2006b)]{rea06b} Read, J. I., Wilkinson, M. I., Evans, N. W., Gilmore, G., \& Kleyna, J. T. 2006b, \mnras, 367, 387

\bibitem[Recchi(2014)]{recc14} Recchi, S. 2014, Advances in Astronomy, 2014, id.750754, 30 pages

\bibitem[Revaz et al.(2009)]{rev09} Revaz, Y., Jablonka, P., Sawala, T., et al. 2009, \aap, 501, 189

\bibitem[Rossi et al.(2008)]{ros08} Rossi, P., Mignone, A., Bodo, G., Massaglia, S., \& Ferrari, A. 2008, \aap, 488, 795

\bibitem[Ruiz et al.(2013)]{rui13} Ruiz, L. O., Falceta-Gon\c{c}alves, D., Lanfranchi, G. A., \& Caproni, A. 2013, \mnras, 429, 1437

\bibitem[Salucci \& Persic(1997)]{sape97} Salucci, P., \& Persic, M. 1997, in ASP Conf. Ser. 117, Dark and Visible
Matter in Galaxies, ed. M. Persic \& P. Salucci (San Francisco: ASP), 1

\bibitem[Schartmann et al.(2010)]{sch10} Schartmann, M., Burkert, A., Krause, M., et al. 2010, \mnras, 403, 1801

\bibitem[Schmidt(1959)]{schm59} Schmidt M., 1959, \apj, 129, 243

\bibitem[Silich \& Tenorio-Tagle(1998)]{site98} Silich, S., \& Tenorio-Tagle, G. 1998, \mnras, 299, 249

\bibitem[Silich \& Tenorio-Tagle(2001)]{site01} Silich, S., \& Tenorio-Tagle, G. 2001, \apj, 552, 91

\bibitem[Simon et al.(2005)]{sim05} Simon, J. D., Bolatto, A. D., Leroy, A., Blitz, L., \& Gates, E. L. 2005, \apj, 621, 757

\bibitem[Simpson et al.(2013)]{sim13} Simpson, C. M., Bryan, G. L., Johnston, K. V. et al. 2013, \mnras, 432, 1989

\bibitem[Spitzer(1982)]{spit82} Spitzer, L., Jr. 1982, \apj, 262, 315

\bibitem[Stinson et al.(2007)]{sti07} Stinson, G. S., Dalcanton, J. J., Quinn, T., Kaufmann, T., \& Wadsley, J. 2007, \apj, 667, 170

\bibitem[Strigari et al.(2007)]{str07} Strigari, L. E., Bullock, J. S., Kaplinghat, M., et al. 2007, \apj, 669, 676

\bibitem[Strigari et al.(2008)]{str08} Strigari, L. E., Bullock, J. S., Kaplinghat, M., et al. 2008, \nat, 454, 1096

\bibitem[Summers(1993)]{summ93} Summers, F. J. 1993, PhD thesis, Univ. California

\bibitem[Sawala et al.(2010)]{saw10} Sawala, T., Scannapieco, C., Maio, U., White, S. 2010, \mnras, 402, 1599

\bibitem[Tesileanu et al.(2008)]{tes08} Tesileanu, O., Mignone, A., \& Massaglia, S. 2008, \aap, 488, 429

\bibitem[van den Berg(1999)]{berg99} van den Bergh, S. 1999, A\&ARv, 9, 273

\bibitem[van den Berg(2007)]{berg07} van den Bergh, S. 2007, The Galaxies of the Local Group (Cambridge: Cambridge Univ. Press)

\bibitem[van den Bosch et al.(2000)]{bos00} van den Bosch, F. C., Robertson, B. E., Dalcanton, J. J., \& de Blok, W. J. G.
2000, \aj, 119, 1579

\bibitem[Vogelsberger et al.(2014)]{vog14} Vogelsberger, M., Genel, S., Springel, V., et al. 2014, \mnras, 444, 1518

\bibitem[Walker et al.(2009)]{wal09} Walker, M. G., Mateo, M., Olszewski, E. W., et al. 2009, \apj, 704, 1274

\bibitem[Walker (2013)]{wal13} Walker, M. G. 2013, in Planets, Stars, and Stellar Systems, Vol. 5: Galactic Structure and Stellar Populations, ed. T. Oswalt \& G. Gilmore (Berlin: Springer), 1039

\bibitem[White \& Rees(1978)]{whre78} White, S. D. M., \& Rees, M. 1978, \mnras, 183, 341

\bibitem[Wiersma, Schaye \& Smith(2009)]{wie09} Wiersma, R. P. C., Schaye, J., \& Smith, B. D. 2009, \mnras, 393, 99

\bibitem[Wilkinson et al.(2004)]{wil04} Wilkinson, M. I., Kleyna, J. T., Evans, N. W., et al. 2004, \apj, 611, 21

\bibitem[Wilson(1955)]{wils55} Wilson, A. G. 1955, \pasp, 67, 27

\bibitem[Wolf et al.(2010)]{wol10} Wolf, J., Martinez, G. D., Bullock, J. S., et al. 2010, \mnras, 406, 1220

\bibitem[Young(1999)]{youn99} Young, L. M. 1999, \aj, 117, 1758

\bibitem[Young(2000)]{youn00} Young, L. M. 2000, \aj, 119, 188

\bibitem[Zingale et al.(2002)]{zin02} Zingale, M., Dursi, L. J., ZuHone, J., et al. 2002, \apjs, 143, 539

\end{thebibliography}
\end{document}